\documentclass[11pt]{amsart}%
\usepackage{amsmath}
\usepackage{amsfonts}
\usepackage{amssymb}
\usepackage{graphicx}%
\providecommand{\U}[1]{\protect\rule{.1in}{.1in}}

\email{chams@aub.edu.lb}
\email{alain@connes.org}
\begin{document}
\vspace{.3cm} \vspace{1cm}

\begin{center}
\baselineskip=16pt \centerline{\Large\bf Spectral Action for Robertson--Walker metrics}
\vspace{2truecm} \centerline{\large\bf Ali H.
Chamseddine$^{1,3,5}$\ , \ Alain Connes$^{2,3,4}$\ \ } \vspace{.5truecm}
\emph{\centerline{$^{1}$Physics Department, American University of Beirut, Lebanon}}
\emph{\centerline{$^{2}$College de France, 3 rue Ulm, F75005, Paris, France}}
\emph{\centerline{$^{3}$I.H.E.S. F-91440 Bures-sur-Yvette, France}}
\emph{\centerline{$^{4}$Department of Mathematics, Vanderbilt University, Nashville, TN 37240 USA}}
\emph{\centerline{$^{5}$LE STUDIUM, Loire Valley Institute for Advanced Studies, Tours and Orleans, France}}
\emph{\centerline{and Laboratoire de Mathematique et Physique Theorique, Universite de Tours, France }}%

\end{center}

\vspace{3cm}

\begin{center}
\textbf{Abstract}
\end{center}

We use the Euler--Maclaurin formula and  the Feynman--Kac formula  to extend our previous method of computation of the spectral action based on the Poisson summation formula. We show how to compute directly the spectral action for  the general case of  Robertson--Walker metrics. We check the terms of the expansion up to $a_6$ against the known universal formulas of Gilkey and compute the expansion up to $a_{10}$ using our direct method.

\tableofcontents


\section{Introduction}

In a previous paper \cite{cc8} we devised a method based on the Poisson summation formula to
compute the spectral action up to arbitrary order for geometric spaces possessing a lot of symmetries such as products of spheres by tori. It is natural to wonder whether this method extends to compute the spectral action for  the general case of  Robertson--Walker metrics. We show in this paper that this is the case. We start by showing in \S \ref{emsect} that for even dimensional spheres one can use the Euler--Maclaurin formula instead of the Poisson summation formula to compute all terms of the asymptotic expansion. Moreover this method is fairly general and applies as soon as the spectral multiplicity is a polynomial and one has a polynomial expression for the squared eigenvalues. To cover the  general case of Robertson--Walker metrics we first reduce in \S \ref{diracsect} the computation of the spectrum of the Dirac operator to a one dimensional problem and then use  the Feynman--Kac formula to compute the spectral action. We first prepare the ground in \S\S \ref{rwsect} and \ref{a6sect}, by performing the computation of the terms up to $a_6$ using the universal formulas of Gilkey. As another preparation we use, in \S \ref{semiclasssect} the semiclassical approximation to get the accurate value of the first term $a_0$. In \S \ref{feynmansect} we explain how to use the Feynman--Kac formula to compute all terms of the expansion, up to surface terms, using the Euler--Maclaurin formula. We check the even terms up to $a_6$ against the universal formulas. This check is quite revealing because of the high degree of complexity of the universal formula of Gilkey for the $a_6$ term. But it leads us to a puzzle since the method yields non vanishing odd terms such as $a_1$, $a_3$ etc. in the expansion. In \S \ref{fulldiracsect} we resolve this puzzle by showing that the operator used in our computation was only similar to the Dirac operator and the natural symmetry of the latter entails the vanishing of the local formulas for the odd terms while justifying the computation of the even ones. We also show that this natural symmetry allows one to use the Poisson summation formula instead of the Euler--Maclaurin formula, and this simplification allows us to compute the local formula up to $a_{10}$ while giving an efficient algorithm to compute terms of arbitrary order. Finally we give in \S \ref{conclsect} some open questions concerning the general terms of the expansion and the conceptual meaning of the use of the Brownian bridge. This paper is a mathematical  preparation for the goal of exploring the cosmological implications of the spectral action principle (\cite{marco1}, \cite{marco}, \cite{mairi}, \cite{mairi1}).

\section{$S^4$, Euler--Maclaurin formula and $\frac{11}{120}$ }\label{emsect}

In this section we explain the relevance of the Euler Maclaurin formula in the computation of the spectral action for spaces whose Dirac spectrum has two properties
\begin{itemize}
  \item The eigenvalues of the Dirac operator $D$ are of the form $\pm h(k)$, where $h(x)^2$ is a polynomial function of $x$.
  \item The multiplicity of $\pm h(k)$ is a polynomial $P(k)$ in $k$.
\end{itemize}
In \cite{cc8} we considered the case of odd spheres with their round  metric and in this case the polynomial $P$ is even so that one can use the Poisson summation formula to compute the spectral action. We also treated the case of the even sphere $S^4$ using the zeta function which is straightforward to compute, and given  by
\begin{equation}
\mathrm{Tr}(|D|^{-s})=\frac 43\left(\zeta(s-3)-\zeta(s-1)\right)\label{4sphere}\,.%
\end{equation}
This function has a  value at $s=0$ given using
\[
\zeta(-3)-\zeta(-1)=\frac{1}{120}+\frac{1}{12}=\frac{11}{120}\,.%
\]
 The same computation gives the full asymptotic expansion of the trace of the heat kernel $\theta(t)={\rm Trace}(e^{-tD^2})$ for the sphere $S^4$, since
$$
    \zeta_{S^4}(s)=\frac{1}{\Gamma(s/2)}\int_0^\infty {\rm Trace}(e^{-tD^2})\,t^{s/2-1}dt
    =\frac{1}{\Gamma(s/2)}\int_0^\infty \theta(t)\,t^{s/2-1}dt
    $$
so that $\zeta_{S^4}(-2k)=(-1)^kk! b_k$ where the $b_k$ are the coefficients of the expansion
$$
    \theta(t)={\rm Trace}(e^{-tD^2})= \frac 23(t^{-2}-t^{-1})+\sum_{k=0}^m b_kt^k+O(t^{m+1})
    $$
     Now recall that the value of the Riemann zeta function at odd negative integers
    is given by
    $$
    \zeta(1-2k)=-\frac{B_{2k}}{2k}
    $$
in terms of the Bernoulli numbers.
Thus one gets for the sphere $S^4$ the heat kernel coefficients
\begin{equation}\label{sphereall}
   b_k=\frac{(-1)^k}{k!}\zeta_{S^4}(-2k)=
    \frac{(-1)^k}{k!}\frac 43\left(\frac{B_{2k+2}}{2k+2}-\frac{B_{2k+4}}{2k+4}\right)
\end{equation}

The Bernoulli numbers are the value at $u=0$ of the Bernoulli polynomials $B_n(u)$ which are defined inductively as follows
\begin{equation*}
    B_0(x)=1\,, \ \ B'_n(x)=nB_{n-1}(x)\,, \ \ \int_0^1B_n(x)dx=0.
\end{equation*}
Equivalently, these polynomials can be introduced using the generating function
\begin{equation}\label{berpo}
   F(u,t)=\frac{t e^{ut}}{e^t-1}=\sum_{n=0}^\infty B_n(u)\frac{t^n}{n!}.
\end{equation}
The presence of the Bernoulli numbers in \eqref{sphereall} can be traced back to the Euler Maclaurin formula
$$
 \sum_{k=a}^bg(k)=\int_a^bg(x)dx +\frac{g(a)+g(b)}{2}+\sum_{j=2}^{m}\frac{B_{j}}{j!}(g^{(j-1)}(b)-
    g^{(j-1)}(a)) - R_m
$$
$$
 R_m=\frac{(-1)^m}{m!}\int_a^b g^{(m)}(x)B_{m}(x-[x])dx
$$
which we shall now use to compute the expansion in the even sphere case. For $S^4$,  the eigenvalues of Dirac are the $\pm k$ with multiplicity $\frac 23(k^3-k)$. This shows that the eigenvalues $0$ and $\pm 1$ are not there (multiplicity zero) and the multiplicity of the eigenvalue $k^2$ of $D^2$ is $\frac 43(k^3-k)$. We first take the test function
$$
f(u)=e^{-tu^2}
$$
We apply the Euler Maclaurin formula  for $a=0$, $b=\infty$ and
$$
g(x)=\frac 43(x^3-x)f(x)=\frac 43(x^3-x)e^{-tx^2}
$$
The Taylor expansion of $g$ at $0$ is given by
\begin{equation}\label{tayl}
g(x)=\frac 43\sum_0^\infty (x^3-x)(-1)^n t^n  x^{2n}/n!=-\frac 43 x+ \frac 43 x^3(1+t)+\ldots
\end{equation}
and one has
$$
g^{(1)}(0)=-\frac 43\,, \ \ \frac{g^{(3)}(0)}{3!}=\frac 43(1+t)\,, \ \ g^{(j-1)}(0)=O(t)\,, \ \  \forall j>4
$$
Thus the only contribution of the term
$$
\sum_{j=2}^{m}\frac{B_{j}}{j!}(g^{(j-1)}(b)-
    g^{(j-1)}(a))=-\sum_{j=2}^{m}\frac{B_{j}}{j!}
    g^{(j-1)}(0)
    $$
    to the constant term in $t$ is of the form
    $$
    -\sum_{j=2}^{4}\frac{B_{j}}{j!}
    g^{(j-1)}(0)=\frac 43\left(\frac{B_{2}}{2}-\frac{B_{4}}{4}\right)=\frac 43\left(\frac{1}{12}+\frac{1}{120}\right)=\frac 43\frac{11}{120}
    $$
The term $\frac{g(a)+g(b)}{2}$ is zero since $g(0)=g(\infty)=0$. The integral
$$
\int_a^bg(x)dx =\frac 43\int_0^\infty (x^3-x)e^{-tx^2}dx =\frac 23(t^{-2}-t^{-1})
$$
does not contribute to the constant term in $t$.

    Let us now use \eqref{tayl} and the Euler Maclaurin formula to compute the full asymptotic expansion in powers of $t$. One has
    $$
    \frac{g^{(2m-1)}(0)}{(2m-1)!}=(-1)^m\frac 43\left(\frac{t^{m-2}}{(m-2)!}+\frac{t^{m-1}}{(m-1)!}\right)
    $$
    The relevant contribution is
    $$
    -\frac{B_{2m}}{2m!}g^{(2m-1)}(0)=-(-1)^m\frac{B_{2m}}{2m}\frac 43\left(\frac{t^{m-2}}{(m-2)!}+\frac{t^{m-1}}{(m-1)!}\right)
    $$
    Thus we get that the coefficient of $t^k/k!$ is given by
    $$
    (-1)^k\frac 43\left(\frac{B_{2k+2}}{2k+2}-\frac{B_{2k+4}}{2k+4}\right)
    $$
       The Euler--Maclaurin formula holds for any smooth function and the only thing one needs to care about is to control the remainder which is of the form
      $$
 R_m=\frac{(-1)^m}{m!}\int_0^\infty g^{(m)}(x)B_{m}(x-[x])dx
$$
and whose size is controlled by an estimate of the form
\begin{equation}\label{controlrem}
|R_m|\leq 2\zeta(m)(2\pi)^{-m}\int_0^\infty|g^{(m)}(x)|dx
\end{equation}
In the case of the spectral action for $S^4$ we apply the Euler--Maclaurin formula to the function
$
g(x)=f(tx^2)(x^3-x)
$
and the derivatives $g^{(m)}(0)$ of this odd function at $x=0$ are of the  form,
\begin{eqnarray}
 g'(0) &=& -f(0) \nonumber \\
 g^{(3)}(0)&=&   6 f(0)-6 t f'(0)                                  \nonumber \\
 g^{(5)}(0)&=&   120 \, t f'(0)-60\, t^2 f''(0)                                  \nonumber \\
 g^{(7)}(0)&=&    2520\, t^2 f''(0)-840\, t^3 f^{(3)}(0)                                 \nonumber \\
 g^{(9)}(0)&=&   60480\, t^3 f^{(3)}(0)-15120 \,t^4 f^{(4)}(0)                                  \nonumber \\
 g^{(11)}(0)&=&    1663200\, t^4 f^{(4)}(0)-332640 \,t^5 f^{(5)}(0)                                 \nonumber \\
 g^{(13)}(0)&=& 51891840 \,t^5 f^{(5)}(0)-8648640\, t^6 f^{(6)}(0)                                    \nonumber
\end{eqnarray}
The Euler--Maclaurin formula then applies directly and gives the expansion
\begin{equation}\label{finesphere}
   \frac 34{\rm Trace}(f(tD^2))=\int_0^\infty f(tx^2)(x^3-x)dx+\frac{11 f(0)}{120}-\frac{31 f'(0) t}{2520}
\end{equation}
$$
+\frac{41 f''(0) t^2}{10080}-\frac{31 f^{(3)}(0) t^3}{15840}+\frac{10331 f^{(4)}(0) t^4}{8648640}-\frac{3421 f^{(5)}(0) t^5}{3931200}+\ldots +R_m
$$
where the remainder $R_m$ is under control. To see this one can use \eqref{controlrem} and the fact that
$$
g^{(m)}(\frac{ u}{\sqrt{t}})=t^{(m-3)/2}P_m(u,t,f^{(k)}(u^2))
$$
where $P_m(u,t,f^{(k)}(u^2))$ is a polynomial in $u$,$t$ and finitely many derivatives of $f$ evaluated at $u^2$. In particular this gives $R_m=O(t^{(m-4)/2})$ when $f$ is smooth with compact support or more generally belongs to the Schwartz space. This implies that when $f$ is a cutoff function, i. e. a smooth function with compact support with all higher derivatives $f^{(k)}(0)=0$
for all $k>0$, the full asymptotic expansion is of the form (up to an overall $\frac 43$)
$$
\frac 34{\rm Trace}(f(tD^2))=\frac{1}{2t^2}\int_0^\infty f(u)udu-\frac{1}{2t}\int_0^\infty f(u)du+\frac{11 f(0)}{120}+O(t^n)\, .
$$
The advantages of this method are that not only one controls the size of the remainder for arbitrary test functions $f$ but also that it can be used  generally for spaces for which one has a polynomial expression $Q(k)$ for the squared eigenvalues of the Dirac operator and a polynomial multiplicity $P(k)$. One applies the same method as above with
$$
g(x)=f(t\,Q(x))P(x).
$$

\section{Dirac operator for Robertson--Walker metrics}

\label{diracsect} The starting point in the evaluation of a spectral action is
the Dirac operator of the relevant geometry. The (Euclidean) Robertson-Walker
metric of dimension $4$ with the symmetry of the round sphere $ S^{3}$ is read from the line element
\begin{equation}
ds^{2}=\ dt^{2}+a^{2}\left(  t\right)  \left(  d\chi^{2}+\sin^{2}\chi\left(
d\theta^{2}+\sin^{2}\theta d\varphi^{2}\right)  \right)  .
\end{equation}
The computation is most efficiently done by using a frame $\theta^{a}$ so that
the line element simplifies to $ds^{2}=\theta^{a}\theta^{a}$ where the flat
indices $a,b,\cdots,$ are contracted with the Euclidean metric $\delta_{ab}.$
Thus the frame is given by
\begin{align}
\theta^{0} &  =dt\\
\theta^{1} &  =a\left(  t\right)  d\chi\\
\theta^{2} &  =a\left(  t\right)  \sin\chi d\theta\\
\theta^{3} &  =a\left(  t\right)  \sin\chi\sin\theta d\varphi
\end{align}
which satisfies%
\begin{align}
d\theta^{0} &  =0\\
d\theta^{1} &  =\frac{a^{\prime}}{a}\theta^{0}\wedge\theta^{1}\\
d\theta^{2} &  =\frac{a^{\prime}}{a}\theta^{0}\wedge\theta^{2}+\frac{\cot\chi
}{a}\theta^{1}\wedge\theta^{2}\\
d\theta^{3} &  =\frac{a^{\prime}}{a}\theta^{0}\wedge\theta^{3}+\frac{\cot\chi
}{a}\theta^{1}\wedge\theta^{3}+\frac{\cot\theta}{a\sin\chi}\theta^{2}%
\wedge\theta^{3}.
\end{align}
where $a^{\prime}$ denotes time derivative. The spin-connection is defined by
\begin{align}
d\theta^{a} &  =\omega^{ab}\wedge\theta^{b}\\
\omega^{ab} &  =\ \theta^{c}\omega_{c}^{\,\,ab}%
\end{align}
The non-vanishing components of $\omega_{abc}$ are
\begin{align}
\omega_{101} &  =\omega_{202}=\omega_{303}=\frac{a^{\prime}}{a}\\
\omega_{212} &  =\omega_{313}=\frac{\cot\chi}{a},\qquad\omega_{323}=\frac
{\cot\theta}{a\sin\chi}.
\end{align}
This shows that the Dirac operator
\begin{align}
D &  =\gamma^{a}e_{a}^{\mu}\frac{\partial}{\partial x^{\mu}}+\frac{1}{4}%
\gamma^{c}\omega_{cab}\gamma^{ab}\\
&  =\gamma^{0}\left(  \frac{\partial}{\partial t}+\frac{3a^{\prime}}%
{2a}\right)  +\frac{1}{a}D_{3}%
\end{align}
where the gamma matrices $\gamma^{a}$ are antihermitian satisfying $\left(
\gamma^{a}\right)  ^{2}=-1,$ and %
\begin{equation}
D_{3}=\gamma^{1}\left(  \frac{\partial}{\partial\chi}+\cot\chi\right)
+\gamma^{2}\frac{1}{\sin\chi}\left(  \frac{\partial}{\partial\theta}+\frac
{1}{2}\cot\theta\right)  +\gamma^{3}\frac{1}{\sin\chi\sin\theta}\frac
{\partial}{\partial\varphi}.
\end{equation}
$D_3$ is directly related to the Dirac operator on $S^{3}$ which is given by
$$
{\rm Dirac}_{S^3}=i\sigma^{1}\left(  \frac{\partial}{\partial\chi}+\cot\chi\right)
+i\sigma^{2}\frac{1}{\sin\chi}\left(  \frac{\partial}{\partial\theta}+\frac
{1}{2}\cot\theta\right)  +i\sigma^{3}\frac{1}{\sin\chi\sin\theta}\frac
{\partial}{\partial\varphi}.
$$
where the $\sigma^j$ are the Pauli matrices. In fact more precisely one has
\begin{equation}\label{squaredir}
\gamma^{0}D_3\sim {\rm Dirac}_{S^3}\oplus -{\rm Dirac}_{S^3}
\end{equation}
using the representation
\begin{equation}
\gamma^{0}=-i\left(
\begin{array}
[c]{cc}%
0 & 1\\
1 & 0
\end{array}
\right)  ,\qquad\gamma^{i}=\left(
\begin{array}
[c]{cc}%
0 & \sigma^{i}\\
-\sigma^{i} & 0
\end{array}
\right)
\end{equation}
Moreover $\gamma^{0}D_3$ has the same square as $D_3$ since $\gamma^{0}$ anti commutes with $D_3$ and has square $-1$.
Thus the square of the Dirac operator is
\begin{equation}\label{diracsquare}
D^{2}=-\left(  \frac{\partial}{\partial t}+\frac{3a^{\prime}}{2a}\right)
^{2}+\frac{1}{a^{2}}(\gamma^{0}D_3)^{2}-\frac{a^{\prime}}{a^{2}}\gamma^{0}D_3%
\end{equation}
One can view the vectors in the Hilbert space of spinors as functions $\xi(t)$ of $t$ with values in
$$
L^2(S^3,\Sigma)\oplus L^2(S^3,\Sigma)
$$
where $\Sigma$ is the spinor bundle on $S^3$ and the square norm of $\xi$ is given by
$$
\int ||\xi(t||^2 a(t)^3 dt
$$
corresponding to the volume form for the Robertson--Walker metric
$$
\sqrt g=a^3 \sin^2\chi\sin\theta
$$
The spectrum of the Dirac operator for the round sphere $S^{d}$ of unit
radius is given by
\begin{equation*}
\mathrm{Spec}(D)=\{\pm(\frac{d}{2}+k)\;|\;k\in{\mathbb{Z}},k\geq 0\} \label{spD}%
\end{equation*}
where the multiplicity of $(\frac{d}{2}+k)$ is equal to $2^{[\frac{d}{2}%
]}{\binom{k+d-1}{k}}$.
For $d=3$ we get the multiplicity $(k+1)(k+2)$ for the eigenvalues $(\frac{3}{2}+k)$, which means the multiplicity $x^2-\frac 14$ for all odd half-integers.
Thus decomposing the vector valued function $\xi(t)$ in terms of the eigenfunctions of the operators $\pm{\rm Dirac}_{S^3}$ for the eigenvalue $\lambda$, one reduces the problem to the direct sum of the one dimensional problems corresponding to the scalar operators in one dimension
given by
\begin{equation}\label{onediracsquare}
-\left(  \frac{\partial}{\partial t}+\frac{3a^{\prime}}{2a}\right)
^{2}+\frac{1}{a^{2}}\lambda^{2}-\frac{a^{\prime}}{a^{2}}\lambda,\ \ \lambda=\pm (\frac 32+n)
\end{equation}
One can move back to the standard measure $dt$ in the variable $t$ by the transformation
\begin{equation}
v\left(  t\right)  =a^{-\frac{3}{2}}u\left(  t\right)
\end{equation}
which gives the direct sum of the operators $H_n^\pm $ where
\begin{equation}
H_n^\pm=-\left(  \frac{d^{2}}{dt^{2}}-\frac{\left(  n+\frac{3}{2}\right)  ^{2}}{a^{2}%
}\pm\frac{\left(  n+\frac{3}{2}\right)  a^{\prime}}{a^{2}}\right)
\end{equation}
which occurs with multiplicity $2(n+1)(n+2)$.
We shall in \S \ref{semiclasssect} and \S \ref{feynmansect} compute the spectral functions of the direct sum of the operators $H_n=H_n^+$ with the multiplicity given by
\begin{equation}
\mu\left(  n\right)  =4(n+1)(n+2)
\end{equation}
The obtained result will be shown to differ by a surface term from the correct result for the original four-dimensional eigenvalue problem, which corresponds to the direct sum of the operators $H_n^+\oplus H_n^-$ with multiplicity $\frac 12 \mu(n)$. This important nuance coming from
the sign in front of the term $\left(  n+\frac{3}{2}\right)  a^{\prime}$  will be dealt with at the beginning of \S \ref{fulldiracsect} using time reversal.

 As a simple example we take the case of the $4$-sphere where $a(t)=\sin(t)$ and $t\in [0,\pi]$. We find that the spectrum of the operator $H_n^\pm$ is given by
$$
{\rm Spec}\, H_n^\pm=\{k^2\mid k\geq n+2\}
$$
We can expand individually the terms $\mathrm{Tr}\left(  e^{-sH_n } \right)$ in the infinite sum:
$$
\mathrm{Tr}\left(  e^{-sH_n } \right)=\sum_{k\geq n+2}e^{-sk^2}=\frac 12\sum_{\mathbb Z} e^{-sk^2}
-\frac 12 -\sum_1^{n+1}e^{-s k^2}\sim\frac{\sqrt{\pi }}{2 \sqrt{s}}
-\frac 12 -\sum_1^{n+1}e^{-s k^2}
$$
$$
=\frac{\sqrt{\pi }}{2 \sqrt{s}}-(n+\frac 32)+ \frac{1}{3} (n+\frac 32) \left(2+3 n+n^2\right) s
+\ldots
$$
This expansion gives a divergent sum for the summation over $n$ and is useless so it is better to use the zeta function. Taking $f(x)=x^{-s/2}$ instead of $f(x)=e^{-sx^2}$ in the above formula we get, using
$$
\sum_0^{k-2}(n+1)(n+2)=\frac{k^3-k}{3}
$$
the correct expression
$$
\mathrm{Tr}(|D|^{-s})={\displaystyle\sum_{n\geq 0}}
\mu(n)\mathrm{Tr}f\left(  H_n  \right)=
{\displaystyle\sum_{n\geq 0}}\mu(n)\sum_{k\geq n+2}k^{-s}=4\sum_{k=2}^\infty\frac{k^3-k}{3}k^{-s}
$$
$$
=\frac 43\left(\zeta(s-3)-\zeta(s-1)\right)
$$

\section{Spectral action for Robertson-Walker metrics}
\label{rwsect}

In this section we give the spectral action for the Robertson-Walker metric 
starting from the universal formula of Gilkey. To do this we will have to
specify all geometric quantities. Starting from
\begin{align}
d\theta^{a} &  =\omega^{ab}\wedge\theta^{b}\\
R^{ab} &  =d\omega^{ab}-\omega^{ac}\wedge\omega^{cb}%
\end{align}
so that the torsion
\begin{equation}
T^{a}=d\theta^{a}-\omega^{ab}\theta^{b}=0
\end{equation}
and $dT^{a}=-R^{ab}\theta^{b}.$ This is consistent with our definition of the
Lorentz covariant derivative
\begin{equation}
D_{\mu}=\partial_{\mu}+\frac{1}{4}\omega_{\mu ab}\gamma_{ab}%
\end{equation}
so that
\begin{align}
\left[  D_{\mu},D_{\nu}\right]   &  =\frac{1}{4}R_{\mu\nu ab}\gamma_{ab}\\
R_{\mu\nu ab} &  =\partial_{\mu}\omega_{\nu ab}-\omega_{\mu ac}\omega_{\nu
cb}-\mu\longleftrightarrow\nu
\end{align}
We also have the relations
\begin{align*}
\omega^{ab} &  =\theta^{c}\omega^{cab}\\
R^{ab} &  =\frac{1}{2}R^{abcd}\theta^{c}\wedge\theta^{d}\\
\theta^{a} &  =e_{\mu}^{a}dx^{\mu}\\
e_{a} &  =e_{a}^{\mu}\partial_{\mu}\\
\theta^{a}\left(  e_{b}\right)   &  =\delta_{b}^{a}%
\end{align*}
Covariant derivatives with respect to curved indices are related to those of
tangent space indices through the equation%
\begin{equation}
\partial_{\mu}e_{\nu a}-\omega_{\mu ab}e_{\nu b}-\Gamma_{\mu\nu}^{\rho}\left(
g\right)  e_{\rho a}=0
\end{equation}
which makes the covariant derivative with respect to tangent space vectors
\begin{equation}
D_{\mu}V_a=\partial_{\mu}V_a -\omega_{\mu ab}V_b
\end{equation}
equivalent to covariant derivatives with respect to vectors over the manifold
\begin{equation}
\nabla_{\mu}V_{\nu}=\partial_{\mu}V_{\nu}-\Gamma_{\mu\nu}^{\rho}V_{\rho}%
\end{equation}
where $V_{\nu}=e_{\nu}^{a}V_{a}.$ Going back to the Robertson-Walker metric we
get the components of the curvatures are
\begin{align*}
R_{01} &  =\frac{a^{\prime\prime}}{a}\theta^{0}\wedge\theta^{1}\\
R_{02} &  =\frac{a^{\prime\prime}}{a}\theta^{0}\wedge\theta^{2}\\
R_{03} &  =\frac{a^{\prime\prime}}{a}\theta^{0}\wedge\theta^{3}\\
R_{12} &  =\frac{\left(  a^{\prime2}-1\right)  }{a^{2}}\theta^{1}\wedge
\theta^{2}\\
R_{13} &  =\frac{\left(  a^{\prime2}-1\right)  }{a^{2}}\theta^{1}\wedge
\theta^{3}\\
R_{23} &  =\frac{\left(  a^{\prime2}-1\right)  }{a^{2}}\theta^{2}\wedge
\theta^{3}%
\end{align*}
which give the relations
\begin{align}
R_{0101} &  =R_{0202}=R_{0303}=\frac{a^{\prime\prime}}{a}\\
R_{1212} &  =R_{1313}=R_{2323}=\frac{\left(  a^{\prime2}-1\right)  }{a^{2}}.
\end{align}
The components of the Ricci tensor are then
\begin{align}
R_{00} &  =R_{0c0c}=3\frac{a^{\prime\prime}}{a}\\
R_{11} &  =R_{1c1c}=\frac{a^{\prime\prime}}{a}+2\frac{\left(  a^{\prime
2}-1\right)  }{a^{2}}=R_{22}=R_{33}%
\end{align}
and finally
\begin{equation}
R=R_{aa}=6\left(  \frac{a^{\prime\prime}}{a}+\frac{\left(  a^{\prime
2}-1\right)  }{a^{2}}\right)  .
\end{equation}
The Weyl tensor defined by
\begin{equation*}
C_{abcd}=R_{abcd}-\frac{1}{2}\left(  \delta_{ac}R_{bd}-\delta_{bc}%
R_{ad}-\delta_{ad}R_{bc}+\delta_{bd}R_{ac}\right)  +\frac{1}{6}\left(
\delta_{ac}\delta_{bd}-\delta_{ad}\delta_{bc}\right)  R
\end{equation*}
vanishes for the Robertson-Walker metric:%
\begin{align}
C_{0101}  & =R_{0101}-\frac{1}{2}\left(  R_{00}+R_{11}\right)  +\frac{1}%
{6}R=0\\
C_{1212}  & =R_{1212}-\frac{1}{2}\left(  R_{11}+R_{22}\right)  +\frac{1}{6}R=0
\end{align}
The Gauss-Bonnet term
$
R^{\ast}R^{\ast}   =\frac{1}{4}\epsilon^{\mu\nu\rho\sigma}\epsilon
_{\alpha\beta\gamma\delta}R_{\mu\nu}^{\quad\alpha\beta}R_{\rho\sigma}%
^{\quad\gamma\delta}
$
 gives
$$
R^{\ast}R^{\ast}=\frac{24 \left(-1+a'(t)^2\right) a''(t)}{a(t)^3}
$$
For an operator of the form%
\begin{equation}
D^{2}=-\left(  g^{\mu\nu}\partial_{\mu}\partial_{\nu}+\mathcal{A}^{\mu
}\partial_{\mu}+B\right)
\end{equation}
the first three Seeley-de Witt coefficients are
\begin{align}
a_{0} &  =\frac{1}{16\pi^{2}}%
{\displaystyle\int}
d^{4}x\sqrt{g}\text{Tr}\left(  1\right)  \\
a_{2} &  =\frac{1}{16\pi^{2}}%
{\displaystyle\int}
d^{4}x\sqrt{g}\text{Tr}\left(  E-\frac{1}{6}R\right)  \\
a_{4} &  =\frac{1}{16\pi^{2}\times360}%
{\displaystyle\int}
d^{4}x\sqrt{g}\text{Tr}\left(   -12\square R+5R^{2}-2R_{\mu\nu}%
^{2}+2R_{\mu\nu\rho\sigma}^{2}  \right.  \nonumber\\
&  \qquad\qquad\qquad\qquad\left.  +60\square E+180E^{2}-60RE+30\Omega_{\mu
\nu}^{2}\right)  \
\end{align}
where%
\begin{align}
\mathbb{\omega}_{\mu} &  =\frac{1}{2}g_{\mu\nu}\left(  \mathcal{A}^{\nu
}+\Gamma^{\nu}\right)  \\
E &  =B-g^{\mu\nu}\left(  \partial_{\mu}\mathbb{\omega}_{\nu}+\mathbb{\omega
}_{\mu}\mathbb{\omega}_{\nu}-\Gamma_{\mu\nu}^{\rho}\mathbb{\omega}_{\rho
}\right)  \\
\Omega_{\mu\nu} &  =\partial_{\mu}\mathbb{\omega}_{\nu}-\partial_{\nu
}\mathbb{\omega}_{\mu}+\left[  \mathbb{\omega}_{\mu},\mathbb{\omega}_{\nu
}\right]
\end{align}
For the above simple Dirac operator we have
\begin{align}
\mathbb{\omega}_{\mu} &  =\frac{1}{4}\omega_{\mu}^{ab}\gamma_{ab}\\
E &  =\frac{1}{4}R\\
\Omega_{\mu\nu} &  =\frac{1}{4}R_{\mu\nu}^{\quad ab}\gamma_{ab}%
\end{align}
We thus get the first two coefficients
\begin{equation}\label{a0}
   a_0=\frac{1}{16\pi^{2}}\times 4\times |S_a^3|=\frac{8\pi^2 a^3}{16\pi^{2}}=\frac 12 a^3
\end{equation}
and
\begin{equation}\label{a2}
    a_2=\frac{1}{16\pi^{2}}\times 4\times |S_a^3|\times \frac{R}{12}=
    \frac{a^3}{4}\left(  \frac{a^{\prime\prime}}{a}+\frac{\left(  a^{\prime
2}-1\right)  }{a^{2}}\right)
\end{equation}

Next in order to compute the coefficient $a_{4}$ we need the geometric
invariants
\begin{equation}
R_{ab}R_{ab}=12\left(  \left(  \frac{a^{\prime\prime}}{a}\right)  ^{2}%
+\frac{\left(  a^{\prime2}-1\right)  ^{2}}{a^{4}}+\frac{a^{\prime\prime
}\left(  a^{\prime2}-1\right)  }{a^{3}}\right)
\end{equation}%
\begin{equation}
R_{abcd}R_{abcd}=12\left(  \left(  \frac{a^{\prime\prime}}{a}\right)
^{2}+\frac{\left(  a^{\prime2}-1\right)  ^{2}}{a^{4}}\right)
\end{equation}
This gives
\begin{equation}
5R^{2}-8R_{ab}R_{ab}-7R_{abcd}R_{abcd}=264\frac{a^{\prime\prime}\left(
a^{^{\prime}2}-1\right)  }{a^{3}}%
\end{equation}
While the surface term comes from $-12\square R+60\square E=3 \square R$ with
\begin{align}
\square R &  =R_{;aa}=D_{a}e_{a}^{0}R^{\prime}=R^{\prime\prime}-\omega
_{aa0}R^{\prime}\nonumber\\
&  =6\left(  \frac{a^{\left(  4\right)  }}{a}+3\frac{a^{\prime}a^{\left(
3\right)  }}{a^{2}}+\frac{\left(  a^{\prime\prime}\right)  ^{2}}{a^{2}}%
-5\frac{a^{\prime2}a^{\prime\prime}}{a^{3}}+2\frac{a^{\prime\prime}}{a^{3}%
}\right)
\end{align}
so that $3 \text{Tr}(1)\square R $ gives a factor $3\times 4\times 6=72$ in front of the 
parenthesis. This gives the final result for $a_{4}:$%
\begin{align}\label{gilka4}
a_{4}  =\frac{2\pi^{2}}{16\pi^{2}\times360}& \int dta^{3}\left(  72\left(
\frac{a^{\left(  4\right)  }}{a}+3\frac{a^{\prime}a^{\left(  3\right)  }%
}{a^{2}}+\frac{\left(  a^{\prime\prime}\right)  ^{2}}{a^{2}}-5\frac
{a^{\prime2}a^{\prime\prime}}{a^{3}}+2\frac{a^{\prime\prime}}{a^{3}}\right)\right.
\nonumber\\\left.+\frac{264}{a^{3}}a^{\prime\prime}\left(  a^{^{\prime}2}-1\right)  \right)
&  =\frac{1}{120}%
{\displaystyle\int}
dt\left(  3a^{2}a^{\left(  4\right)  }+9aa^{\prime}a^{\left(  3\right)
}+3aa^{^{\prime\prime}2}-4a^{\prime2}a^{\prime\prime}-5a^{\prime\prime
}\right)  \ .
\end{align}
We note that this is a linear combination of the total derivative $\square R$
and the Gauss-Bonnet topological combination. Thus $a_{4}$ does not
contribute to the dynamics of the spectral action. The above formula fits with the general 
formula 
$$
a_4=\frac{1}{16\pi^{2}}\frac{1}{360}%
{\displaystyle\int}
d^{4}x\sqrt{g}\left(  -18C_{\mu\nu\rho\sigma}^{2}+11R^{\ast}R^{\ast}+12\square R\right)
$$
since the Weyl tensor vanishes.
\section{Calculation of $a_{6}$}

\label{a6sect}

To calculate the next term in the spectral action we start with Gilkey's
formula. This is fairly complicated, we thus partition the expression into
four blocks. The first block depends quadratically on covariant derivatives of
the curvature tensor and its contractions:
\begin{align*}
&  \frac{1}{16\pi^{2}\times7!}\int d^{4}x\sqrt{g}\mathrm{Tr}\left(
-18\square^{2}R+17R_{;a}R_{;a}-2R_{ab;c}R_{ab;c}-4R_{ab;c}R_{bc;a}\ \right.
\\
&  \qquad\qquad\left.  +9R_{abcd;e}R_{abcd;e}+28R\square R-8R_{ab}\square
R_{ab}+24R_{ab}R_{bc;ac}+12R_{abcd}\square R_{abcd}\right)  \
\end{align*}
The second block depends cubically on the curvature $R_{abcd}$ and its
contractions but not its derivatives:
\begin{align*}
& \frac{1}{16\pi^{2}\times 9  \times 7!}\int d^{4}x\sqrt{g}%
\mathrm{Tr}\left(  -35R^{3}+42RR_{ab}R_{ab}-42RR_{abcd}R_{abcd}+208R_{ab}%
R_{bc}R_{ca}\ \right.  \\
&  \qquad\left.  -192R_{ab}R_{cd}R_{acbd}+48R_{ab}R_{acde}%
R_{bcde}-44R_{abcd}R_{abef}R_{cdef}-80R_{abcd}R_{aecf}R_{bedf}\right)
\end{align*}
The third  block depends on the curvatures $\Omega_{ab}$ and their derivatives
as well as the curvatures $R_{abcd}$ and their contractions:
\begin{align*}
&  \frac{\mathrm{\ 1}}{16\pi^{2}\times360}\int d^{4}x\sqrt{g}\mathrm{Tr}%
\left(  8\Omega_{ab;c}\Omega_{ab;c}+2\Omega_{ab;b}\Omega_{ac;c}+12\Omega
_{ab}\square\Omega_{ab}\ \right.  \\
&  \qquad\qquad\qquad\left.  -12\Omega_{ab}\Omega_{bc}\Omega_{ca}%
-6R_{abcd}\Omega_{ab}\Omega_{cd}+4R_{ab}\Omega_{ac}\Omega_{bc}-5R\Omega
_{ab}\Omega_{ab}\right)
\end{align*}
The fourth block depends on $E,$ $\Omega_{ab}$, $R_{abcd}$ and its
contractions
\begin{align*}
&  \frac{\mathrm{\ 1}}{16\pi^{2}\times360}\int d^{4}x\sqrt{g}\mathrm{Tr}%
\left(  6\square^{2}E+60E\square E+30E_{;a}E_{;a}+60E^{3}+30E\Omega_{ab}%
\Omega_{ab}-10R\square E\right.  \\
&  \qquad\left.  -4R_{ab}E_{;ab}-12R_{;a}E_{;a}-30RE^{2}-12\square
RE+5R^{2}E-2R_{ab}R_{ab}E+2R_{abcd}R_{abcd}E\right)  .
\end{align*}
In our case
$
E    =\frac{1}{4}R.1$ and
$\Omega_{\mu\nu}    =\frac{1}{4}R_{\mu\nu ab}\gamma_{ab}$.
Calculating the traces and using the identities
\begin{align*}
\mathrm{Tr}\left(  \gamma_{ab}\gamma_{cd}\right)   &  =-4\left(  \delta
_{ac}\delta_{bd}-\delta_{ad}\delta_{bc}\right)  \\
\mathrm{Tr}\left(  \gamma_{ab}\gamma_{cd}\gamma_{ef}\right)   &  =4\left[
\left(  \ \delta_{bc}\left(  \delta_{ae}\delta_{df}-\delta_{af}\delta
_{de}\right)  \ -a\longleftrightarrow b\right)  -c\longleftrightarrow
d\right]  \
\end{align*}
we obtain the following table of conversion:
$$
\begin{array}{ccc}
 \Omega_{ab}^2\to -\frac{1}{2}R_{abcd}^2\,, & \Omega
_{ab}\square\Omega_{ab}\to -\frac{1}{2}R_{abcd}\square R_{abcd}\,, & \Omega_{ab;c}^2\to -\frac{1}{2}R_{abcd;e}^2 \\
\Omega_{ac}\Omega_{bc}\to  -\frac{1}{2} R_{acde}R_{bcde} \,, & \Omega_{ab;b}\Omega_{ac;c}\to 
   -\frac{1}{2}R_{abcd;b}R_{aecd;e}\,, & \Omega_{ab}\Omega_{cd}\to  -\frac{1}{2}R_{abef}R_{cdef}\\
 & \Omega_{ab}\Omega_{bc}\Omega_{ca}\to -\frac{1}{2}R_{abcd}R_{aecf}R_{bedf} &   
\end{array}
$$
Note that all other terms have an additional factor $\mathrm{Tr}(1)=4$ in the computation of their trace.
This enables us to list all terms that appear in $a_{6}:$%
\begin{align}
&  \frac{\mathrm{\ }1}{16\pi^{2}}%
{\displaystyle\int}
d^{4}x\sqrt{g}\left(  \frac{1}{420}\square^{2}R+\frac{1}{1008}R_{;a}%
R_{;a}-\frac{1}{630}R_{ab;c}R_{ab;c}-\frac{1}{315}R_{ab;c}R_{ac;b}\right.
\nonumber\\
&  \qquad\qquad\qquad-\frac{1}{252}R_{abcd;e}R_{abcd;e}+\frac{1}{360}R\square
R-\frac{2}{315}R_{ab}\square R_{ab}+\frac{2}{105}R_{ab}R_{bc;ac}\nonumber\\
&  \qquad\qquad\qquad-\frac{1}{140}R_{abcd}\square R_{abcd}+\frac{1}%
{2592}R^{3}-\frac{1}{540}RR_{ab}R_{ab}-\frac{7}{4320}RR_{abcd}R_{abcd}%
\nonumber\\
&  \qquad+\frac{52}{2835}R_{ab}R_{bc}R_{ca}-\frac{16}{945}R_{ab}R_{cd}%
R_{acbd}-\frac{1}{756}R_{ab}R_{acde}R_{bcde}+\frac{101}{22680}R_{abcd}%
R_{abef}R_{cdef}\nonumber\\
&  \qquad\qquad\left.  \ -\frac{1}{90}R_{ab}R_{;ab}-\frac{1}{360}%
R_{abcd;b}R_{aecd;e}+\frac{109}{11\,340}R_{abcd}R_{aecf}R_{bedf}\right)
\end{align}

We see that some of these terms are not so straightforward to evaluate. We
note the following%
\begin{align}
R_{;a} &  =\delta_{a0}R^{\prime}\\
R_{ab;c} &  =e_{c}R_{ab}-\omega_{cad}R_{db}-\omega_{cbd}R_{ad}%
\end{align}
thus since the $R_{ab}$ are dependent only on $t,$ and $\omega_{0ab}=0,$ we
then have
\begin{align}
R_{ab;0} &  =R_{ab}^{\prime}\\
R_{01;1} &  =\frac{a^{\prime}}{a}\left(  R_{00}-R_{11}\right)  =R_{02;2}%
=R_{03;3}%
\end{align}
Notice that this last relation shows that one must be very careful in working
with covariant derivatives which arise because of the split between space and
time coordinates.  Complications start with second derivatives:%
\begin{align*}
\square R_{ab} &  =D_{c}\left(  D_{c}R_{ab}\right)  \\
&  =e_{c}\left(  e_{c}R_{ab}-\omega_{cad}R_{db}-\omega_{cbd}R_{ad}\right)
-\omega_{ccd}\left(  e_{d}R_{ab}-\omega_{dae}R_{eb}-\omega_{dbe}R_{ae}\right)
\\
&  -\omega_{cad}\left(  e_{c}R_{db}-\omega_{cde}R_{eb}-\omega_{cbe}%
R_{de}\right)  -\omega_{cbd}\left(  e_{c}R_{ad}-\omega_{cae}R_{ed}%
-\omega_{cde}R_{ae}\right)
\end{align*}
More complicated is
\begin{align*}
\square R_{abcd} &  =D_{e}\left(  D_{e}R_{abcd}\right)  \\
&  =e_{e}\left(  D_{e}R_{abcd}\right)  -\omega_{eef}D_{f}R_{abcd}-\omega
_{eaf}D_{e}R_{ebcd}\\
&  -\omega_{ebf}D_{e}R_{afcd}-\omega_{ecf}D_{e}R_{abfd}-\omega_{edf}%
D_{e}R_{abcf}%
\end{align*}
where
\[
D_{e}R_{abcd}=e_{e}R_{abcd}-\omega_{eaf}R_{fbcd}-\omega_{ebf}R_{afcd}%
-\omega_{ecf}R_{abfd}-\omega_{edf}R_{abcf}%
\]
 We note that all the components of the tensors such as $$
 R\,,\ R_{;a}\,,\ R_{;ab}\,,\ R_{ab}\,, \ R_{ab;c}\,,\, \ R_{ab;cd}\,,\, \ R_{abcd}\,,\, \ R_{abcd;e}\,,\, \ R_{abcd;ef}$$ are only functions of the variable $t$ and not of the other variables $\chi,\theta, \phi$ even though the variables $\chi,\theta$ appear explicitly in some components of the connection $\omega_{abc}$.
 We omit the lengthy steps, and we just list the final expressions.
We start by the scalar curvature and the fact that the Laplacian on functions $f(t)$
of $t$ alone is given by $\square f=\frac{3 a'(t) f'(t)}{a(t)}+f''(t)$, thus
\begin{equation}\label{scalaaa}
    \square^2R=\left(  R^{^{\prime\prime}}%
+3\frac{a^{\prime}}{a}R^{\prime}\right)  ^{^{\prime\prime}}+3\frac{a^{\prime}%
}{a}\ \left(  R^{^{\prime\prime}}+3\frac{a^{\prime}}{a}R^{\prime}\right)
^{^{\prime}}
\end{equation}
We also have easily $R_{;a}R_{;a}   =\left(  R^{\prime}\right)  ^{2}$. For the next two terms we use
\begin{align}
R_{ab;c}R_{ab;c} &  =\left(  R_{00}^{\prime}\right)  ^{2}+3\left(
R_{11}^{\prime}\right)  ^{2}+6\left(  \frac{a^{\prime}}{a}\right)  ^{2}\left(
R_{00}-R_{11}\right)  ^{2}\\
R_{ab;c}R_{ac;b} &  =\left(  R_{00}^{\prime}\right)  ^{2}+3\left(
\frac{a^{\prime}}{a}\right)  ^{2}\left(  R_{00}-R_{11}\right)  ^{2}%
+6\frac{a^{\prime}}{a}R_{11}^{\prime}\left(  R_{00}-R_{11}\right) 
\end{align}
 Then for the term in $R_{abcd;e}^2$ we get 
 \begin{equation*}
    R_{abcd;e}R_{abcd;e} =12\left[  \left(  R_{0101}^{\prime}\right)
^{2}+\left(  R_{1212}^{\prime}\right)  ^{2}+4\left(  \frac{a^{\prime}}%
{a}\right)  ^{2}\left(  R_{0101}-R_{1212}\right)  ^{2}\right] 
 \end{equation*}
 The term in $R\square R$ is straightforward.
 For the next term in $R_{ab}\square R_{ab}$  we use
 \begin{equation}\label{riccis}
 R_{ab}\square R_{ab}  =R_{00}\square R_{00}+3R_{11}\square R_{11}
 \end{equation}
where moreover
\begin{equation}
\square R_{00}=R_{00}^{^{\prime\prime}}+3\left(  \frac{a^{\prime}}{a}\right)
R_{00}^{\prime}+6\left(  \frac{a^{\prime}}{a}\right)  ^{2}\left(
R_{11}-R_{00}\right)
\end{equation}%
and
\begin{equation}
\square R_{11}=R_{11}^{^{\prime\prime}}+3\left(  \frac{a^{\prime}}{a}\right)
R_{11}^{\prime}+2\left(  \frac{a^{\prime}}{a}\right)  ^{2}\left(
R_{00}-R_{11}\right)  =\square R_{22}=\square R_{33}%
\end{equation}
The next term is in $R_{ab}R_{bc;ac}$ and is given by
\begin{align}
R_{ab}R_{bc;ac}= &  R_{00}\left(  R_{00}^{\prime\prime}+3\frac{a^{\prime}}%
{a}\left(  R_{00}^{\prime}-R_{11}^{\prime}\right)  \ -3\left(  \frac
{a^{\prime}}{a}\right)  ^{2}\left(  R_{00}-R_{11}\right)  \right)  \nonumber\\
&  +3R_{11}\left(  \left(  \frac{a^{\prime\prime}}{a}+3\frac{a^{\prime2}%
}{a^{2}}\right)  \left(  R_{00}-R_{11}\right)  +\frac{a^{\prime}}{a}%
R_{00}^{\prime}\right)
\end{align}
Next for the term involving the full Riemann tensor $R_{abcd}\square R_{abcd}$ one has
\begin{equation}\label{fullrr}
 R_{abcd}\square R_{abcd}   =12\left[  R_{0101}\square R_{0101}+R_{1212}%
\square R_{1212}\right]
\end{equation}
where
\begin{align}
\square R_{0101} &  =R_{0101}^{\prime\prime}+3\left(  \frac{a^{\prime}}%
{a}\right)  R_{0101}^{\prime}+4\left(  \frac{a^{\prime}}{a}\right)
^{2}\left(  R_{1212}-R_{0101}\right)  \\
&  =\square R_{0202}=\square R_{0303}\nonumber
\end{align}%
and
\begin{align}
\square R_{1212} &  =R_{1212}^{\prime\prime}+3\left(  \frac{a^{\prime}}%
{a}\right)  R_{1212}^{\prime}+4\left(  \frac{a^{\prime}}{a}\right)
^{2}\left(  R_{0101}-R_{1212}\right)  \\
&  =\square R_{1313}=\square R_{2323}\nonumber
\end{align}
The term in $R^3$ is straightforward. One has for the next term
\begin{equation}\label{richsqu}
   RR_{ab}R_{ab} =R\left(  R_{00}^{2}+3R_{11}%
^{2}\right)
\end{equation}
Similarly for the next term, $RR_{abcd}R_{abcd}$ involving the full Riemann tensor, we get
\begin{equation}\label{rriiee}
    RR_{abcd}R_{abcd}=12R\left(  R_{0101}^{2}+R_{1212}^{2}\right)
\end{equation}

We now evaluate the three next terms (and remember that $R_{22}=R_{11}$)
\begin{align}
R_{ab}R_{bc}R_{ca} &  =R_{00}^{3}+3R_{11}^{3}\\
R_{ab}R_{cd}R_{acbd} &  =6R_{00}R_{11}R_{0101}+6R_{11}R_{22}R_{1212}\\
R_{ab}R_{acde}R_{bcde} &  =6\left(  R_{00}+R_{11}\right)  R_{0101}%
^{2}+12R_{11}R_{1212}^{2}
\end{align}
The next term is cubic in the Riemann tensor and given by 
$R_{abcd}R_{abef}R_{cdef}$. One gets 
\begin{equation}\label{cubic1}
    R_{abcd}R_{abef}R_{cdef}==24\left(  R_{0101}^{3}+R_{1212}^{3}\right) 
\end{equation}
For the next one $R_{ab}R_{;ab}$ one gets
\begin{equation}
R_{ab}R_{;ab}   =R_{00}R^{\prime\prime}+3\frac{a^{\prime}}{a}R_{11}R^{\prime
}
\end{equation}
We still have one more term involving first derivatives of the Riemann tensor
\begin{equation}\label{derivterm}
R_{abcd;b}R_{aecd;e}  =6\left(  R_{0101}^{\prime}+2\frac{a^{\prime}}%
{a}\left(  R_{0101}-R_{1212}\right)  \right)  ^{2}
\end{equation}
Finally the last term is
\begin{equation}
R_{abcd}R_{aecf}R_{bedf}   =6\left(  R_{1212}^{3}+3R_{1212}R_{0101}%
^{2}\right)
\end{equation}
We are now ready to collect all the terms in one expression%
\begin{align}
&  \frac{1}{8}%
{\displaystyle\int}
dt\,a^{3}\left[  \frac{1}{420}\left(  \left(  R^{^{\prime\prime}}%
+3\frac{a^{\prime}}{a}R^{\prime}\right)  ^{^{\prime\prime}}+3\frac{a^{\prime}%
}{a}\ \left(  R^{^{\prime\prime}}+3\frac{a^{\prime}}{a}R^{\prime}\right)
^{^{\prime}}\right)  +\frac{1}{1008}\left(  R^{\prime}\right)  ^{2}\right.
\nonumber\\
&  \,\qquad\qquad-\frac{1}{630}\left(  \left(  R_{00}^{\prime}\right)
^{2}+3\left(  R_{11}^{\prime}\right)  ^{2}+6\left(  \frac{a^{\prime}}%
{a}\right)  ^{2}\left(  R_{00}-R_{11}\right)  ^{2}\right)  \nonumber\\
&  \qquad\qquad-\frac{1}{315}\left(  \left(  R_{00}^{\prime}\right)
^{2}+3\left(  \frac{a^{\prime}}{a}\right)  ^{2}\left(  R_{00}-R_{11}\right)
^{2}+6\frac{a^{\prime}}{a}R_{11}^{\prime}\left(  R_{00}-R_{11}\right)
\right)  \nonumber\\
&  \qquad\qquad-\frac{1}{21}\left[  \left(  R_{0101}^{\prime}\right)
^{2}+\left(  R_{1212}^{\prime}\right)  ^{2}+4\left(  \frac{a^{\prime}}%
{a}\right)  ^{2}\left(  R_{0101}-R_{1212}\right)  ^{2}\right]  \nonumber\\
&  \qquad\qquad+\frac{1}{360}R\left(  R^{^{\prime\prime}}+3\frac{a^{\prime}%
}{a}R^{\prime}\right)  \nonumber\\
&  \qquad\qquad-\frac{2}{315}\left[  R_{00}\left(  R_{00}^{^{\prime\prime}%
}+3\frac{a^{\prime}}{a}R_{00}^{\prime}+6\frac{a^{\prime2}}{a^{2}}\left(
R_{11}-R_{00}\right)  \right)  \right.  \nonumber\\
&  \qquad\qquad\qquad\qquad\left.  +3R_{11}\left(  R_{11}^{^{\prime\prime}%
}+3\frac{a^{\prime}}{a}R_{11}^{\prime}+2\frac{a^{\prime2}}{a^{2}}\left(
R_{00}-R_{11}\right)  \right)  \right]  \nonumber\\
& +\ \frac{2}{105}\left[  R_{00}\left(  R_{00}^{\prime\prime
}+3\frac{a^{\prime}}{a}\left(  R_{00}^{\prime}-R_{11}^{\prime}\right)
\ -3\left(  \frac{a^{\prime}}{a}\right)  ^{2}\left(  R_{00}-R_{11}\right)
\right)  \right. \nonumber \\
&  \qquad\qquad\left.  +3R_{11}\left(  \left(  \frac{a^{\prime\prime}}%
{a}+3\frac{a^{\prime2}}{a^{2}}\right)  \left(  R_{00}-R_{11}\right)
+\frac{a^{\prime}}{a}R_{00}^{\prime}\right)  \right]  \nonumber\\ 
&  \qquad\qquad-\frac{3}{35}\left[  R_{0101}\left(  R_{0101}^{\prime\prime
}+3\frac{a^{\prime}}{a}R_{0101}^{\prime}+4\frac{a^{\prime2}}{a^{2}}\left(
R_{1212}-R_{0101}\right)  \right)  \ \right.  \nonumber\\
&  \qquad\qquad\qquad\left.  +R_{1212}\ \left(  R_{1212}^{\prime\prime}%
+3\frac{a^{\prime}}{a}R_{1212}^{\prime}+4\frac{a^{\prime2}}{a^{2}}\left(
R_{0101}-R_{1212}\right)  \right)  \right]  \ \nonumber\\
&  \qquad\ +\frac{1}{2592}R^{3}-\frac{1}{\ 540}R\left(  R_{00}^{2}+3R_{11}%
^{2}\right)  -\frac{7\ }{\ 4320}12R\left(  R_{0101}^{2}+R_{1212}^{2}\right)
\nonumber\\
&  \qquad+\frac{52\ }{\ 2835}\left(  R_{00}^{3}+3R_{11}^{3}\right)  -\frac
{32}{315}\left(  R_{00}R_{11}R_{0101}+R_{11}^{2}R_{1212}\right)  \nonumber\\
&  \qquad-\frac{1}{126}\left(  \left(  R_{00}+R_{11}\right)  R_{0101}%
^{2}+2R_{11}R_{1212}^{2}\right)  +\frac{101}{22680}24\left(  R_{0101}%
^{3}+R_{1212}^{3}\right)  \nonumber
\end{align}
\begin{align*}
&  \qquad-\frac{1}{90}\left(  R_{00}R^{\prime\prime}+3\frac{a^{\prime}}%
{a}R_{11}R^{\prime}\right)  -\frac{1}{60}\left(  R_{0101}^{\prime}%
+2\frac{a^{\prime}}{a}\left(  R_{0101}-R_{1212}\right)  \right)
^{2}\nonumber\\
&  \qquad\qquad\qquad\left.  +\frac{109}{1890}\left(  R_{1212}^{3}%
+3R_{1212}R_{0101}^{2}\right)  \right]
\end{align*}
As an elementary test of this we evaluate the $a_{6}$ coefficient for a sphere
$S^{4}$ where $a=\sin t,$ which implies
\[
R=-12,\qquad R_{0101}=R_{1212}=-1,\qquad R_{00}=R_{11}=-3
\]
which gives the coefficient to be
\[
\ \frac{31}{2520}%
\]
and thus
\[
a_{6}=\frac{31}{2520}%
{\displaystyle\int\limits_{0}^{\pi}}
\sin^{3}tdt=\frac{31}{1890}%
\]
which agrees with the expansion given in \eqref{finesphere}.

Substituting the expressions of
$R,R_{00},R_{11},R_{0101},R_{1212}$ into the above formula gives
\begin{align}\label{gilka6}
a_6 &  =-\frac{a'(t)^2 a''(t)}{240 a(t)^2}-\frac{a'(t)^4 a''(t)}{84 a(t)^2}+\frac{a''(t)^2}{120 a(t)}+\frac{a'(t)^2 a''(t)^2}{21 a(t)}-\frac{1}{90} a''(t)^3\\
& +\frac{a'(t) a^{(3)}(t)}{240 a(t)}+\frac{a'(t)^3 a^{(3)}(t)}{84 a(t)}-\frac{1}{20} a'(t) a''(t) a^{(3)}(t)-\frac{a(t) a^{(3)}(t)^2}{1680}-\frac{1}{240} a^{(4)}(t)\nonumber\\
& -\frac{1}{120} a'(t)^2 a^{(4)}(t) +\frac{a(t) a''(t) a^{(4)}(t)}{840}+\frac{1}{140} a(t) a'(t) a^{(5)}(t)+\frac{a(t)^2 a^{(6)}(t)}{560}\nonumber%
\end{align}
This expression is quite elaborate and it will be a major test for our direct method
of computation which we develop below, to check that we obtain the same result for $a_6$. This will be done in \S \ref{feynmansect}. The above equality together with \eqref{a0}, \eqref{a2}, and the computation of $a_4$ determine the spectral action
\begin{equation}
I=\mathrm{Tr}\,f\left(   D^{2}/M_{\mathrm{Pl}}^{2}\right)
=M_{\mathrm{Pl}}^{4}f_{4}a_{0}+M_{\mathrm{Pl}}^{2}f_{2}a_{2}+f_{0}%
a_{4}+M_{\mathrm{Pl}}^{-2}f_{-2}a_{6}
\end{equation}
up to order $\frac{1}{M_{\mathrm{Pl}}^{2}}$
where
\begin{equation}
f_{4}=%
\int_0^\infty
uf\left(  u\right)  du,\,\  f_{2}=%
\int_0^\infty
f\left(  u\right)  du,\  f_{0}=f\left(  0\right)  ,\  f_{-2}%
=-f^{\prime}\left(  0\right)
\end{equation}
As $a_{4}$ does not contribute to the dynamical equation for $a\left(
t\right)  $, $a_{6}$ will have non-trivial effects on the standard model of
cosmology.

\section{Semi-classical approximation and $a_0$}\label{semiclasssect}

As shown in \S \ref{diracsect}, the square of the Dirac operator  for the Robertson--Walker metric associated to the function $a(t)$ is intimately related  to the direct sum of the  operators $H_n$ given by
\begin{align}
&H_n= -\frac{d^{2}}{dt^{2}}+V_{n}\left(  t\right)  \\
V_{n}\left(  t\right)    & =\frac{\left(  n+\frac{3}{2}\right)  }{a^{2}%
}\left(  \left(  n+\frac{3}{2}\right)-a^{\prime}  \right)
\end{align}
where $H_n$ occurs with the multiplicity $\mu(n)=4(n+1)(n+2)$ for all integers $n\geq 0$.
This means that the spectral action corresponds to the following sum
\begin{equation}%
\mathrm{Tr}f(D^2)\sim
{\displaystyle\sum_{n\geq 0}}
\mu(n)\mathrm{Tr}f\left(  H_n  \right)
\end{equation}
We take the functions $f(u)=e^{-su}$ and we look at the expansion
when $s\to 0$. At the semi-classical level the operator is
$$
H_n=p^2+V_n(t)
$$
and in this section we shall investigate what happens if we just use (for $f(u)=e^{-su}$)
the first approximation
to $\mathrm{Tr}f\left(  H_n  \right)$ given by the semi-classical expression
$$
\mathrm{Tr}f\left(  H_n  \right)\sim_{\rm sc} \frac{1}{2\pi}\int e^{-s(p^2+V_n(t))}dp dt
$$
where the integration is over $p\in \mathbb R$ and $t$.  One has
$$
\frac{1}{2\pi}\int e^{-s(p^2+V_n(t))}dp dt=\frac{\sqrt{\pi }}{\sqrt{s}}
\frac{1}{2\pi}\int e^{-sV_n(t)} dt
$$
and thus the semi-classical approximation  is given locally in $t$ by the sum
$$
\sum_{n\geq 0}
\mu(n)\mathrm{Tr}f\left(  H_n  \right)\sim_{\rm sc}\frac{1}{2\sqrt{\pi s}}
\sum_{n\geq 0}
\mu(n)e^{-sV_n(t)}
$$
We can use the Euler-Maclaurin formula to evaluate the summation in $n$ and the first contribution is  an integral. As a function of $n$ we are dealing with
$$
\mu(n)e^{-sV_n(t)}=4(n+1)(n+2)e^{s \frac{\left(  n+\frac{3}{2}\right)  }{a^{2}%
}\left(  a^{\prime}-\left(  n+\frac{3}{2}\right)  \right)} =h_s(n+\frac{3}{2})
$$
where
$$
h_s(x)=(4x^2-1)e^{-s \frac{x^2 }{a^{2}}+s \frac{x a^{\prime}}{a^{2}}
}
$$
Next one has
\begin{equation}\label{erfint}
\int_{\frac 32}^\infty (4x^2-1) e^{u(b-x) x} dx=\frac{1}{2 u^{3/2}}e^{\frac{3}{4} (-3+2 b) u}\ \ \times
\end{equation}
 $$
  \left(2 (3+b) \sqrt{u}+e^{\frac{1}{4} (-3+b)^2 u} \sqrt{\pi } \left(2+\left(-1+b^2\right) u\right) \left(1+\text{Erf}\left[\frac{1}{2} (-3+b) \sqrt{u}\right]\right)\right)
$$
 where the error function is
 $$
 \text{Erf}(z)=\frac{2}{\sqrt{\pi }}\int _0^ze^{-v^2}dv
 $$
 and its expansion near $0$ is
 $$
 \frac{2 z}{\sqrt{\pi }}-\frac{2 z^3}{3 \sqrt{\pi }}+\frac{z^5}{5 \sqrt{\pi }}-\frac{z^7}{21 \sqrt{\pi }}+O[z]^9
 $$
Thus taking $u=s a^{-2}$, $b=a'$, one obtains the expansion of the coefficient of $dt$ after multiplication by  $\frac{1}{2\sqrt{\pi s}}$ which gives up to $O[s]^{1/2}$,
$$\frac{a^3}{2 s^2}+\frac{a^2 a'}{ \sqrt{\pi } s^{3/2}}+\frac{a \left(-2+3 \left(a'\right)^2\right)}{8 s}+\frac{-18-3 a'+4 \left(a'\right)^3}{12 \sqrt{\pi } \sqrt{s}}+\frac{\left(a'\right)^2 \left(-4+5 \left(a'\right)^2\right)}{64 a}
$$
One needs to add the other terms of the Euler-Maclaurin formula, which are
$$
\frac 12 h_s(\frac 32)-\sum_{j=2}^{m}\frac{B_{j}}{j!}
    h_s^{(j-1)}(\frac 32)
$$
The  derivative $h_s^{(a)}$ is of the order of $s^\alpha$ where $\alpha$ is the largest integer $\alpha<a/2$. Thus for the expansion up to $O[s]^{1/2}$ one can take only the terms corresponding to
$$
\frac 12 h_s(\frac 32)-\frac{B_{2}}{2!}
    h_s^{(1)}(\frac 32)=\frac 12 h_s(\frac 32)-\frac{1}{12}h_s^{(1)}(\frac 32)
$$
since $h_s^{(3)}$ is of order $s$. The computation of these terms gives, after multiplication by  $\frac{1}{2\sqrt{\pi s}}$,
$$
\frac{3}{2 \sqrt{\pi } \sqrt{s}}
$$
which cancels the term independent of $a'$ in $\frac{-18-3 a'+4 \left(a'\right)^3}{12 \sqrt{\pi } \sqrt{s}}$ and thus the asymptotic expansion from the semi-classical formula gives up to $O[s]^{1/2}$,
$$
\frac{a^3}{2 s^2}+\frac{a^2 a'}{\sqrt{\pi } s^{3/2}}+\frac{a \left(-2+3 \left(a'\right)^2\right)}{8 s}+\frac{-3 a'+4 \left(a'\right)^3}{12 \sqrt{\pi } \sqrt{s}}+\frac{\left(a'\right)^2 \left(-4+5 \left(a'\right)^2\right)}{64 a}
$$
This formula is not nice at all since
\begin{itemize}
  \item It only involves $a$ and $a'$ and hence cannot reproduce the curvature terms.
  \item It has non vanishing terms in odd powers of $\sqrt s$.
  \item The coefficient of $1/\sqrt{s}$ is not a total derivative.
\end{itemize}
This is of course to be expected since the approximation uses the wrong assumption that $p$ and the function $a(t)$ commute. This assumption only holds
 in the case where $a$ is constant, in that case one drops all terms in $a'$ and only two terms remain:
$$
\frac{a^3}{2 s^2}-\frac{a }{4 s}
$$
which agrees with  \cite{cc8} thm. 3. In general the only term we can trust is the leading term $a_0=\frac{a^3}{2}$ which fits with the general formula
$$
a_0= \frac{\text{Tr}(1)}{16\pi^{2}}%
{\displaystyle\int}
\sqrt{g}d^{4}x
$$
since $\text{Tr}(1)=4$ and the volume of the three sphere of radius $a$ is $2\pi^2 a^3$.

\section{Feynman Kac formula and up to $a_6$ with surface terms}\label{feynmansect}

In this section we use the Feynman Kac formula to compute the spectral action.
We use the following formula (see \cite{Sim} Theorem 6.6 page 54) for the local expression of the trace of $e^{-sH_n}$,
$$
e^{-2s(-\frac 12\partial_t^2+V)}(t,t)=\frac{1}{2\sqrt{\pi s}}\int{\rm exp}\left(-2s\int_0^1V(t+\sqrt{2s}\,\alpha(u))\,du\right)D[\alpha]
$$
where  $\alpha$ is the Brownian bridge (see \cite{Sim} Definition page 40) which is a Gaussian random variable with covariance given by
$$
E(\alpha(u)\alpha(v))=u(1-v)
$$
for $u\leq v$. In our case the potential $V$ is given by $-\frac 12\partial_t^2+V=\frac 12 H_n$
$$
V(t)=\frac 12 V_n(t)
$$
and the above formula becomes
$$
e^{-sH_n}(t,t)=\frac{1}{2\sqrt{\pi s}}\int{\rm exp}\left(-s\int_0^1V_n(t+\sqrt{2s}\,\alpha(v))\,dv\right)D[\alpha]
$$
The technique then is that we evaluate the sum with multiplicity included,
$$
\sum_{n=0}^\infty\mu(n)e^{-sH_n}(t,t)
$$
by using the Euler-Maclaurin formula, which means replacing the discrete index $n$ by the continuous variable $x=n+\frac 32$ and the sum over $n$ by
$$
\int_{\frac 32}^\infty k_s(x)dx+\frac{1}{2} k_s \left(\frac{3}{2}\right)-\frac{k'_s\left(\frac{3}{2}\right)}{12}+\frac{k_s''\left(\frac{3}{2}\right)
}{720}-\frac{k_s^{(4)}\left(\frac{3}{2}\right)}{30240}+\ldots
$$
where the function $k_s$ is given by
$$
k_s(x)=(4x^2-1)\frac{1}{2\sqrt{\pi s}}\int e^{u (b-x) x}D[\alpha]
$$
where
$$
u=s\int_0^1a^{-2}(t+\sqrt{2s}\,\alpha(v))\,dv
$$
and
$$
ub=s\int_0^1a'a^{-2}(t+\sqrt{2s}\,\alpha(v))\,dv
$$
so that
$$
b=\int_0^1a'a^{-2}(t+\sqrt{2s}\,\alpha(v))\,dv/\left(\int_0^1a^{-2}(t+\sqrt{2s}\,\alpha(v))\,dv\right)
$$
To obtain these terms we have used $x=n+\frac 32$ and
$$
-s\int_0^1V_n(t+\sqrt{2s}\,\alpha(v))\,dv=u(b-x)x
$$
We  use the Taylor expansion of $a^{-2}$ and of $a'a^{-2}$ for the expressions of $u$ and $b$ to get an asymptotic expansion when $s\to 0$. The terms of the expansion are coming from the general formula
$$
\int_0^1 F(t+\sqrt{2s}\,\alpha(v))\,dv=F(t)+\sum \frac{F^{(k)}(t)}{k!}(\sqrt{2s})^kx_k(\alpha)
$$
where
$$
x_k(\alpha)=\int_0^1\alpha(v)^k\,dv
$$
We get the following expansions
$$
u=\frac{ s}{a(t)^2}-\frac{2 \left(\sqrt{2} a'(t)\right) x_1(\alpha) s^{3/2}}{a(t)^3}-\frac{2 \left(-3   a'(t)^2+a(t)   a''(t)\right)x_2(\alpha) s^2}{a(t)^4}+O[s]^3
$$
and
$$
b=a'(t)+\sqrt{2} a''(t) x_1(\alpha)\sqrt{s}$$
$$+\frac{\left(4 x_1(\alpha)^2 a'(t) a''(t)-4 x_2(\alpha) a'(t) a''(t)+a(t) x_2(\alpha) a^{(3)}(t)\right) s}{a(t)}+O[s]^{3/2}
$$
We are first dealing with the integral term in the Euler-Maclaurin formula. Thus we can use the formula \eqref{erfint}. The leading term is unchanged and given by
$
a(t)^3/(2 s^2)
$ as above.
When we use the additional terms for $u$ and $b$ we find that the term in $\frac 1s$ in the expansion gets new terms depending on $\alpha$, and becomes
$$
\frac{a(t) \left(-2+3 a'(t)^2\right)}{8 s}+ \frac{15 a(t) a'(t)^2}{2 s}x_1(\alpha)^2
+\frac{3 a(t) \left(-3 a'(t)^2+a(t) a''(t)\right)}{2 s}x_2(\alpha)
$$
Performing the Gaussian integral one gets
$$
\int x_1(\alpha)^2D[\alpha]=\frac{1}{12}\,, \ \ \int x_2(\alpha)D[\alpha]=\frac{1}{6}
$$
and  the term in $\frac 1s$ thus gives
$$
\frac{a(t) \left(-2+3 a'(t)^2\right)}{8 s}+ \frac{15 a(t) a'(t)^2}{2 s}\frac{1}{12}
+\frac{3 a(t) \left(-3 a'(t)^2+a(t) a''(t)\right)}{2 s}\frac{1}{6}
$$
$$
=\frac{a(t) \left(-1+a'(t)^2+a(t) a''(t)\right)}{4 s}
$$
This agrees with the
spectral action which gives for the $\frac 1s$ term
\[
a_{2}=\frac{1}{4\pi^{2}}
{\displaystyle\int}  \frac{R}{12}
\sqrt{g}d^{4}x%
\]
while
\[
R=6\left(  \frac{a''}{a}+\frac{a^{\prime2}}{a^{2}}-\frac{1}{a^{2}}\right)
\]
which is negative for the sphere and gives (using $|S_a^3|=2\pi^2 a^3$)
\[
a_{2}=\frac{1}{4}%
{\displaystyle\int}
dta^{3}\left(  \frac{a''}{a}+\frac{a^{\prime2}}{a^{2}}-\frac{1}{a^{2}}\right)
\]
We now look at the $a_3$ term, including the corrections it gives
$$
a_3=\frac{1}{6} \left(a'(t) \left(-3+4 a(t) a''(t)\right)+2 a(t)^2 a^{(3)}(t)\right)
$$
and one finds that it is the total derivative of the following expression which vanishes
at both ends of the time interval
$$
-\frac{a(t)}{2}+\frac{1}{3} a(t)^2 a''(t)
$$
The next term is the $a_4$ term, one performs the computation in the same way as above and obtains
$$
a_4=\frac{1}{120} \left(-\left(5+4 a'(t)^2\right) a''(t)+3 a(t) a''(t)^2+3 a(t) \left(3 a'(t) a^{(3)}(t)+a(t) a^{(4)}(t)\right)\right)
$$
This term agrees with \eqref{gilka4} and is very interesting because it does not vanish for the sphere case where it gives
$$a_4({\rm sphere})=\frac{11 \sin^{3}t}{120}\,, \ \ \int_0^\pi a_4({\rm sphere})dt
=\frac 43\times\frac{11}{120}$$
but it is a ``topological" term since it is the derivative of the following expression
$$
-\frac{1}{24} a'(t)-\frac{7 a'(t)^3}{360}+\frac{1}{40} a(t) a'(t) a''(t)+\frac{1}{40} a(t)^2 a^{(3)}(t)
$$
In this expression the last two terms vanish at the end points of the time interval, but not the first two. In fact the variation of the first two across the interval will not be zero in general except when $a'(t)$ vanishes at the boundary which is the case of $S^1\times S^3$. In the sphere case the derivative at the end points is $\pm 1$ and this corresponds to smoothing out the conical
singularity at the boundary.

The computation of $a_6$ is more complicated. To do the computation one needs to compute the integrals of polynomials in the $x_j(\alpha)$ under the Gaussian measure $D[\alpha]$ in order to obtain the coefficients. We list in the appendix the table of the integrals which are needed to compute up to $a_{10}$. It gives
$$
a_6=-\frac{a'(t)^2 a''(t)}{240 a(t)^2}-\frac{a'(t)^4 a''(t)}{84 a(t)^2}+\frac{a''(t)^2}{120 a(t)}+\frac{a'(t)^2 a''(t)^2}{21 a(t)}-\frac{1}{90} a''(t)^3+\frac{a'(t) a^{(3)}(t)}{240 a(t)}$$ $$+\frac{a'(t)^3 a^{(3)}(t)}{84 a(t)}-\frac{1}{20} a'(t) a''(t) a^{(3)}(t)-\frac{a(t) a^{(3)}(t)^2}{1680}-\frac{1}{240} a^{(4)}(t)-\frac{1}{120} a'(t)^2 a^{(4)}(t)$$ $$+\frac{a(t) a''(t) a^{(4)}(t)}{840}+\frac{1}{140} a(t) a'(t) a^{(5)}(t)+\frac{a(t)^2 a^{(6)}(t)}{560}
$$
One checks that it agrees with the computation \eqref{gilka6} using the Gilkey universal formula.
It is important to see why the discrete terms in the Euler-Maclaurin formula do not contribute to $a_6$. In fact the direct computation gives (up to a factor of $4$)
$$
\frac{99 x_1(\alpha) a'(t)}{20 \sqrt{2} a(t)^3}-\frac{29 \sqrt{2} x_1(\alpha) a'(t)^2}{15 a(t)^3}+\frac{29 x_1(\alpha) a''(t)}{15 \sqrt{2} a(t)^2}
$$
and such terms disappear after the integration in $D(\alpha)$.
In the general case the terms $a_6$ and $a_8$, since they have the denominator $a(t)$ to some power, will be singular at $t=0$, but this singularity vanishes provided that
$a^{(2)}(0)=0$ in the case of $a_6$. These conditions are verified for the sphere. In fact the smoothness of the
metric at the pole $t=0$ is equivalent to the condition
$$
a'(0)=\pm 1\,, \ \ a^{(2n)}(0)=0, \ \ \forall n.
$$
The results of this section leave us with a puzzle  since we get the correct values for the even terms but the method yields non vanishing odd terms such as $a_1$, $a_3$ etc. They give surface terms whose integral vanishes but they seem to contradict the vanishing of the {\em local} trace for the odd terms.

\section{Full Dirac operator, Poisson summation and up to $a_{10}$}\label{fulldiracsect}

In this section we resolve the above puzzle by showing that the operator used above, namely the direct sum of the operators $H_n=H_n^+$ with multiplicity $\mu(n)$, admits the same even terms in its spectral expansion as  the Dirac operator and the natural symmetry of the latter entails the vanishing of the local formulas for the odd terms while justifying the computation of the even ones. We also show that this natural symmetry allows one to use the Poisson summation formula instead of the Euler--Maclaurin formula, and this simplification allows us to compute the local formula up to $a_{10}$ while giving an algorithm to compute terms of arbitrary order.

We have seen in \S \ref{diracsect} that the square of the Dirac operator   is
\begin{equation*}
D^{2}=\bigoplus \frac 12 \mu(n)\left(H_n^+\oplus H_n^-\right)
\end{equation*}
Now observe that the operator $H_n^-$ can be viewed as the ``time reversal" of $H_n=H_n^+$
\begin{align}
&H_n^T= -\frac{d^{2}}{dt^{2}}+V_{n}^T\left(  t\right)  \\
V_{n}^T\left(  t\right)    & =\frac{\left(  n+\frac{3}{2}\right)  }{a^{2}%
}\left( \left(  n+\frac{3}{2}\right)+ a^{\prime} \right)
\end{align}
and this implies that the spectral asymptotics for the direct sum of the $H_n^-$
with multiplicity $\mu(n)$ are obtained from the above ones simply by replacing the derivatives $a^{(k)}$ by $(-1)^k a^{(k)}$. Thus no new computation is needed for the first terms and one checks that, for the Dirac operator, the non vanishing odd terms such as $a_1$, $a_3$ etc. cancel out, while the even terms remain unchanged. We shall in fact go further and show how to use this added symmetry to simplify the above computations trading the Euler--Maclaurin formula for the Poisson summation formula.
We consider the function
$$
f_s(x)=(x^2-\frac 14)e^{u(b-x)x}
$$
and the relevant sum is now
\begin{equation}\label{relsum}
    \sum_{-\infty}^\infty f_s(n+\frac 12)
\end{equation}
Using the Poisson summation formula it is very well approximated by
$$
\int_{-\infty}^\infty f_s(x+\frac 12)dx=\frac{e^{\frac{b^2 u}{4}} \sqrt{\pi } \left(2+\left(-1+b^2\right) u\right)}{4 u^{3/2}}
$$
There is an overall factor of $2$ coming from spinors and one multiplies by $\frac{1}{2\sqrt{\pi s}}$ and then integrates in $D[\alpha]$ as above with
$$
u=s\int_0^1a^{-2}(t+\sqrt{2s}\,\alpha(v))\,dv
$$
and
$$
ub=s\int_0^1a'a^{-2}(t+\sqrt{2s}\,\alpha(v))\,dv
$$
so that
$$
b=\int_0^1a'a^{-2}(t+\sqrt{2s}\,\alpha(v))\,dv/\left(\int_0^1a^{-2}(t+\sqrt{2s}\,\alpha(v))\,dv\right)
$$
Thus the relevant expression is
\begin{equation}\label{finalform}
\frac 14\int\frac{e^{\frac{b^2 u}{4}}  \left(2+\left(-1+b^2\right) u\right)}{\sqrt{s}\,u^{3/2}}D[\alpha]
\end{equation}
One then repeats the same computation as above and finds that all the terms $a_j$ for $j$ odd vanish, while they are unchanged for even $j$. For instance when we compute $a_4$ we find the following coefficients as a polynomial expression in the $x(j)$
$$
\begin{array}{c}
 x(1)^4\to \frac{315 a'(t)^4}{4 a(t)} \\
 x(1)^2 x(2)\to -\frac{315 a'(t)^4}{2 a(t)}+\frac{105}{2} a'(t)^2 a''(t) \\
 x(1)^2\to -\frac{3 a'(t)^2}{4 a(t)}+\frac{9 a'(t)^4}{8 a(t)}+\frac{9}{2} a'(t)^2 a''(t)+\frac{3}{4} a(t) a''(t)^2 \\
 x(1) x(3)\to \frac{60 a'(t)^4}{a(t)}-45 a'(t)^2 a''(t)+5 a(t) a'(t) a^{(3)}(t) \\
 x(2)^2\to \frac{135 a'(t)^4}{4 a(t)}-\frac{45}{2} a'(t)^2 a''(t)+\frac{15}{4} a(t) a''(t)^2 \\
 x(2)\to \frac{3 a'(t)^2}{4 a(t)}-\frac{9 a'(t)^4}{8 a(t)}-\frac{a''(t)}{4}-\frac{21}{8} a'(t)^2 a''(t)+\frac{3}{4} a(t) a'(t) a^{(3)}(t) \\
 x(4)\to -\frac{15 a'(t)^4}{a(t)}+18 a'(t)^2 a''(t)-\frac{9}{4} a(t) a''(t)^2-3 a(t) a'(t) a^{(3)}(t)+\frac{1}{4} a(t)^2 a^{(4)}(t) \\
 1\to -\frac{a'(t)^2}{16 a(t)}+\frac{5 a'(t)^4}{64 a(t)}
\end{array}
$$
 We then replace the monomials in the $x(j)$ by their integral under $D[\alpha]$ computed using integration by parts under the Gaussian measure and gathered in  the table of the Appendix.

 The main result then is that this technique gives the full expansion of the spectral action to arbitrary order. The proof is the same as in \cite{cc8} using the fact that the remainder in
the Poisson summation formula has flat Taylor expansion.

For the $a_8$ term one obtains the following expression
$$
a_8=-\frac{1}{10080 a(t)^4}P(a)
$$
\begin{math}
P(a)=108 a'(t)^6 a''(t)-108 a(t) a'(t)^5 a^{(3)}(t)+a'(t)^4 (27 a''(t)-588 a(t) a''(t)^2\\+60 a(t)^2 a^{(4)}(t))-3 a(t) a'(t)^3 ((9-256 a(t) a''(t)) a^{(3)}(t)+8 a(t)^2 a^{(5)}(t))+3 a(t) a'(t)^2 (-29 a''(t)^2+267 a(t) a''(t)^3 -104 a(t)^2 a''(t) a^{(4)}(t)+a(t) (-72 a(t) a^{(3)}(t)^2+5 a^{(4)}(t)+3 a(t)^2 a^{(6)}(t)))+a(t)^2 a'(t) (-819 a(t) a''(t)^2 a^{(3)}(t)+6 a''(t) (17 a^{(3)}(t)+13 a(t)^2 a^{(5)}(t))+a(t) (132 a(t) a^{(3)}(t) a^{(4)}(t)-6 a^{(5)}(t)-5 a(t)^2 a^{(7)}(t)))+a(t)^2 (43 a''(t)^3-114 a(t) a''(t)^4+69 a(t)^2 a''(t)^2 a^{(4)}(t)+2 a(t) a''(t) (45 a(t) a^{(3)}(t)^2-18 a^{(4)}(t)+a(t)^2 a^{(6)}(t))+a(t) (-24 a^{(3)}(t)^2+13 a(t)^2 a^{(4)}(t)^2+16 a(t)^2 a^{(3)}(t) a^{(5)}(t)+3 a(t) a^{(6)}(t)-a(t)^3 a^{(8)}(t)))
\end{math}

\medskip
and for the sphere $S^4$ it simplifies to
$$a_8({\rm sphere})=\frac{41\,\sin^{3}t}{10080}\,, \ \ \int_0^\pi a_8({\rm sphere})dt
=\frac 43\times\frac{41}{10080}$$
which agrees with the direct computation \eqref{finesphere} of \S \ref{emsect}.

With this method we can now reach $a_{10}$ which is given by

\bigskip

\begin{math}
a_{10}=\frac{1}{665280  a(t)^6}(-11700 a'(t)^8 a''(t)+11700 a(t) a'(t)^7 a^{(3)}(t)\\ +3 a'(t)^6 (5 a''(t) (-165+5096 a(t) a''(t))-2046 a(t)^2 a^{(4)}(t))\\+3 a(t) a'(t)^5 ((825-34628 a(t) a''(t)) a^{(3)}(t)+746 a(t)^2 a^{(5)}(t))\\+3 a(t) a'(t)^4 (3476 a''(t)^2-54054 a(t) a''(t)^3+14440 a(t)^2 a''(t) a^{(4)}(t)\\+a(t) (10448 a(t) a^{(3)}(t)^2-429 a^{(4)}(t)-217 a(t)^2 a^{(6)}(t)))\\+3 a(t)^2 a'(t)^3 (78902 a(t) a''(t)^2 a^{(3)}(t)-2 a''(t) (2222 a^{(3)}(t)+2127 a(t)^2 a^{(5)}(t))\\+a(t) (-7992 a(t) a^{(3)}(t) a^{(4)}(t)+154 a^{(5)}(t)+55 a(t)^2 a^{(7)}(t)))\\+a(t)^2 a'(t)^2 (-11880 a''(t)^3+111378 a(t) a''(t)^4-68664 a(t)^2 a''(t)^2 a^{(4)}(t)\\+33 a(t) (113 a^{(3)}(t)^2+124 a(t)^2 a^{(4)}(t)^2+192 a(t)^2 a^{(3)}(t) a^{(5)}(t)-4 a(t) a^{(6)}(t))\\+3 a(t) a''(t) (-31973 a(t) a^{(3)}(t)^2+1738 a^{(4)}(t)+968 a(t)^2 a^{(6)}(t))-43 a(t)^4 a^{(8)}(t))\\+a(t)^3 a'(t) (-117600 a(t) a''(t)^3 a^{(3)}(t)+66 a''(t)^2 (211 a^{(3)}(t)+172 a(t)^2 a^{(5)}(t))\\-2 a(t) a''(t) (-19701 a(t) a^{(3)}(t) a^{(4)}(t)+693 a^{(5)}(t)+238 a(t)^2 a^{(7)}(t))\\+a(t) (-2640 a^{(3)}(t) a^{(4)}(t)-2 a(t)^2 (778 a^{(4)}(t) a^{(5)}(t)+537 a^{(3)}(t) a^{(6)}(t))\\+33 a(t) (271 a^{(3)}(t)^3+a^{(7)}(t))+18 a(t)^3 a^{(9)}(t)))+a(t)^3 (2354 a''(t)^4\\-3 a(t) a''(t) (3446 a''(t)^4+1243 a^{(3)}(t)^2+924 a''(t) a^{(4)}(t))+66 a(t)^2 (331 a''(t)^2 a^{(3)}(t)^2\\+160 a''(t)^3 a^{(4)}(t)+6 a^{(4)}(t)^2+9 a^{(3)}(t) a^{(5)}(t)+4 a''(t) a^{(6)}(t))-a(t)^3 (331 a^{(3)}(t)^2 a^{(4)}(t)\\+482 a''(t) a^{(4)}(t)^2+1110 a''(t) a^{(3)}(t) a^{(5)}(t)+448 a''(t)^2 a^{(6)}(t)+11 a^{(8)}(t))\\-3 a(t)^4 (74 a^{(5)}(t)^2+116 a^{(4)}(t) a^{(6)}(t)+49 a^{(3)}(t) a^{(7)}(t)+6 a''(t) a^{(8)}(t))\\+3 a(t)^5 a^{(10)}(t)))
\end{math}

\medskip

When we evaluate this expression for $a(t)=\sin(t)$ we obtain for the sphere $S^4$ the result
$$a_{10}({\rm sphere})=\frac{31\, \sin^{3}t}{15840}\,, \ \ \int_0^\pi a_8({\rm sphere})dt
=\frac 43\times\frac{31}{15840}$$
which agrees with the direct computation \eqref{finesphere} of \S \ref{emsect}. There is another very useful test of the above computations which is to compute the divergence at $t=0$. The test is that this divergence should cancel under the hypothesis that $a^{(2n)}=0$ and $a'(0)=1$. For instance for the coefficient $a_{10}$ above, one finds that the numerator admits, under the hypothesis that $a'(0)=1$ and $a^{(2n)}=0$ for $n\leq 4$, the expansion

\medskip

\begin{math}
\frac{1155}{128} a^{(10)}(0) t^8+\frac{22}{35}(2625 a^{(3)}(0)^5-12712 a^{(3)}(0)^3 a^{(5)}(0) +6804 a^{(3)}(0) a^{(5)}(0)^2\\ +2592 a^{(3)}(0)^2 a^{(7)}(0)-1152 a^{(5)}(0) a^{(7)}(0)-240 a^{(3)}(0) a^{(9)}(0)) t^9+O(t)^{10}
\end{math}

\medskip
which is of order $9$ in $t$ provided $a^{(10)}(0)=0$ and thus compensates the denominator $a(t)^9$ in the formula for the local coefficient of the volume form $a(t)^3 dt$.


\section{Conclusion}\label{conclsect}

We have developed in this paper a direct method for the computation of the spectral action for Robertson--Walker metrics in Euclidean formulation. We have shown that our method, based on the Euler--Maclaurin formula combined with the Feynman--Kac formula, gives the same result for the local terms of the expansion as the Gilkey formulas up to $a_6$. We have computed the full expansion up to $a_{10}$ and checked its accuracy by performing concrete tests such as the direct comparison in the case of the $4$-sphere.  We have shown also how to control the remainder in the asymptotic expansion which only gives an approximate expression of the spectral action neglecting the role of the instanton contributions. In fact besides giving the above explicit formulas our results suggest the following questions:
\vspace{.05in}

$\bullet$~Check the agreement between the above formulas for $a_8$ and $a_{10}$ and the universal formulas of \cite{Amst}, \cite{Avra} and \cite{Van}.

\vspace{.05in}

$\bullet$~Show that the term $a_{2n}$ of the asymptotic expansion of the spectral action for Robertson--Walker metrics is of the form $P_n(a,\ldots,a^{(2n)})/a^{2n-4}$ where $P_n$ is a polynomial with rational coefficients, and compute $P_n$.
 
 \vspace{.05in}

$\bullet$~Is there a conceptual physical meaning for the role of the integration over the Brownian bridge as a fluctuation of the time variable in passing from the semiclassical approximation to the full spectral action?
 \vspace{.05in}

\section{Appendix}

We give in this appendix the table of integrals of polynomials in the $x_j(\alpha)$ under the Gaussian measure $D[\alpha]$ corresponding to the Brownian bridge. The computations are straightforward  since they consist in integrating by parts  under a Gaussian but time consuming
so that it is useful to get the assistance of a computer. We list below the integrals needed to compute up to $a_{10}$.
The meaning of an arrow such as $x_1^2 x_3^2\to \frac{659}{33600}$ is
$$
\int
 x_1(\alpha)^2 x_3(\alpha)^2    D[\alpha] =\frac{659}{33600}
$$
where $\alpha(t)$ is the Brownian bridge and
$$
x_k(\alpha)=\int_0^1\alpha(v)^kdv
$$

$$
\begin{array}{cccc}
 1\to 1 & x_1^2\to \frac{1}{12} & x_1^4\to \frac{1}{48} & x_1^6\to \frac{5}{576} \\
 x_1^8\to \frac{35}{6912} & x_1^{10}\to \frac{35}{9216} & x_2\to \frac{1}{6} & x_1^2 x_2\to \frac{11}{360} \\
 x_1^4 x_2\to \frac{17}{1440} & x_1^6 x_2\to \frac{23}{3456} & x_1^8 x_2\to \frac{203}{41472} & x_2^2\to \frac{1}{20} \\
 x_1^2 x_2^2\to \frac{83}{5040} & x_1^4 x_2^2\to \frac{893}{100800} & x_1^6 x_2^2\to \frac{57}{8960} & x_2^3\to \frac{61}{2520} \\
 x_1^2 x_2^3\to \frac{259}{21600} & x_1^4 x_2^3\to \frac{187}{22400} & x_2^4\to \frac{1261}{75600} & x_1^2 x_2^4\to \frac{4097}{369600} \\
 x_2^5\to \frac{79}{5280} & x_1 x_3\to \frac{1}{20} & x_1^3 x_3\to \frac{1}{56} & x_1^5 x_3\to \frac{13}{1344} \\
 x_1^7 x_3\to \frac{1}{144} & x_1 x_2 x_3\to \frac{43}{1680} & x_1^3 x_2 x_3\to \frac{439}{33600} & x_1^5 x_2 x_3\to \frac{7}{768} \\
 x_1 x_2^2 x_3\to \frac{227}{12600} & x_1^3 x_2^2 x_3\to \frac{4471}{369600} & x_1 x_2^3 x_3\to \frac{2579}{158400} & x_3^2\to \frac{23}{560} \\
 x_1^2 x_3^2\to \frac{659}{33600} & x_1^4 x_3^2\to \frac{1303}{98560} & x_2 x_3^2\to \frac{31}{1120} & x_1^2 x_2 x_3^2\to \frac{1877}{105600} \\
 x_2^2 x_3^2\to \frac{3851}{158400} & x_1 x_3^3\to \frac{467}{17600} & x_4\to \frac{1}{10} & x_1^2 x_4\to \frac{5}{168} \\
 x_1^4 x_4\to \frac{17}{1120} & x_1^6 x_4\to \frac{85}{8064} & x_2 x_4\to \frac{19}{420} & x_1^2 x_2 x_4\to \frac{527}{25200} \\
 x_1^4 x_2 x_4\to \frac{69}{4928} & x_2^2 x_4\to \frac{5}{168} & x_1^2 x_2^2 x_4\to \frac{1493}{79200} & x_2^3 x_4\to \frac{7171}{277200} \\
 x_1 x_3 x_4\to \frac{9}{280} & x_1^3 x_3 x_4\to \frac{127}{6160} & x_1 x_2 x_3 x_4\to \frac{15629}{554400} & x_3^2 x_4\to \frac{7939}{184800} \\
 x_4^2\to \frac{23}{420} & x_1^2 x_4^2\to \frac{1817}{55440} & x_2 x_4^2\to \frac{6353}{138600} & x_1 x_5\to \frac{3}{56} \\
 x_1^3 x_5\to \frac{17}{672} & x_1^5 x_5\to \frac{1495}{88704} & x_1 x_2 x_5\to \frac{1}{28} & x_1^3 x_2 x_5\to \frac{839}{36960} \\
 x_1 x_2^2 x_5\to \frac{383}{12320} & x_3 x_5\to \frac{19}{336} & x_1^2 x_3 x_5\to \frac{7543}{221760} & x_2 x_3 x_5\to \frac{117}{2464} \\
 x_1 x_4 x_5\to \frac{3067}{55440} & x_5^2\to \frac{355}{3696} & x_6\to \frac{3}{28} & x_1^2 x_6\to \frac{5}{112} \\
 x_1^4 x_6\to \frac{139}{4928} & x_2 x_6\to \frac{11}{168} & x_1^2 x_2 x_6\to \frac{4283}{110880} & x_2^2 x_6\to \frac{3001}{55440} \\
 x_1 x_3 x_6\to \frac{13}{220} & x_4 x_6\to \frac{61}{616} & x_1 x_7\to \frac{1}{12} & x_1^3 x_7\to \frac{13}{264} \\
 x_1 x_2 x_7\to \frac{109}{1584} & x_3 x_7\to \frac{19}{176} & x_8\to \frac{1}{6} & x_1^2 x_8\to \frac{71}{792}\\
 x_2 x_8\to \frac{17}{132}& x_1 x_9\to \frac{15}{88}& x_{10}\to \frac{15}{44}
\end{array}
$$

\section*{Acknowledgment}

The research of A. H. C. is supported in part by the National Science
Foundation under Grant No. Phys-0854779.


\begin{thebibliography}{99}                                                                                               %

\bibitem {Amst} P. Amsterdamski, A. Berkin, and D. O'Connor, ``$b_8$ Hamidew coefficient for
a scalar field", Classical Quantum Gravity 6 (1989), 1981--1991.



\bibitem {Avra} I. G. Avramidi, ``The covariant technique for the calculation of the heat
kernel asymptotic expansion", Phys. Lett. B 238 (1990), 92--97.


\bibitem {BG1}T. Branson and P. Gilkey, ``Residues for the eta function for an
operator of Dirac type with local boundary conditions", \textit{Diff. Geom.
Appl. }\textbf{2 }(1992) 249.

\bibitem {BG2}T. Branson and P. Gilkey, ``Residues of the eta function for an
operator of Dirac type", \textit{Journal of Functional Analysis }\textbf{108
}(1992) 47.

\bibitem {BGV}T. Branson, P. Gilkey and D. Vassilevich, ``Vacuum expectation
value asmyptotics for second order differential operators on manifolds with
boundary", \textit{J. Math. Phys. }\textbf{39 }(1998) 1040.

\bibitem{CH} R. Camporesi and A. Higuchi, \textquotedblleft
On the eigenfunctions of the Dirac operator on spheres and real hyperbolic spaces",
\textit{J. Geom. Phys. }\textbf{20}1 (1996).

\bibitem {ACAC}Ali H. Chamseddine and Alain Connes,  \textquotedblleft
Universal Formula for Noncommutative Geometry Actions: Unification of Gravity
and the Standard Model", \textit{Phys. Rev. Lett. }\textbf{77 }4868 (1996);
"The Spectral Action Principle" \textit{Comm. Math. Phys. }\textbf{186 }731 (1997).

\bibitem {ACM}A.~H.~Chamseddine, A.~Connes and M.~Marcolli, \emph{Gravity and
the standard model with neutrino mixing}
Adv.\ Theor.\ Math.\ Phys.\ \textbf{11}, 991 (2007) [arXiv:hep-th/0610241].

\bibitem {scale}Ali H. Chamseddine and Alain Connes, \textquotedblleft Scale
Invariance in the Spectral Action" \textit{J. Math. Phys. }\textbf{47
}063504\textbf{ }(2006).

\bibitem {Beggar}A.~H.~Chamseddine and A.~Connes, \emph{Conceptual Explanation
for the Algebra in the Noncommutative Approach to the Standard Model}
Phys.\ Rev.\ Lett.\ \textbf{99}, 191601 (2007) [arXiv:0706.3690 [hep-th]]

\bibitem {Why}A.~H.~Chamseddine and A.~Connes, \emph{Why the Standard Model}
J.\ Geom.\ Phys.\ \textbf{58}, 38 (2008) [arXiv:0706.3688 [hep-th]].

\bibitem {quantumactions}Ali H. Chamseddine and Alain Connes,
\textquotedblleft Quantum Gravity Boundary Terms from Spectral Action",
\textit{Phys. Rev. Lett. }\textbf{99 }071302 (2007), arXiv:0705.1786.

\bibitem {cc8}
  A.~Connes and A.~H.~Chamseddine,
  \emph{The uncanny precision of the spectral action}. Comm. Math. Phys. 293 (2010), no. 3, 867--897


\bibitem {Gilkeyb1}P. Gilkey ``Invariance Theory, the heat equation and the
Atiyah-Singer Index theorem", CRC press, second edition.

\bibitem {Gilkeyb2}P. Gilkey ``Asymptotic Formulae in Spectral Geometry", CRC
press, 2004.

 \bibitem {marco1} M. Marcolli and E. Pierpaoli {\em Early Universe models from Noncommutative Geometry} arXiv:0908.3683.



\bibitem {marco} M. Marcolli, E. Pierpaoli and K. Teh {\em The spectral action and cosmic topology } arXiv:1005.2256.


\bibitem {MTW}C. Misner K. Thorne and J. Wheeler "Gravitation" section 21.4-21.8

\bibitem {mairi} W. Nelson and M. Sakellariadou {\em Cosmology and the Noncommutative approach to the Standard Model} arXiv:0812.1657.

\bibitem {mairi1} W. Nelson and M. Sakellariadou {\em    Natural inflation mechanism in asymptotic noncommutative geometry} arXiv:0903.1520.

\bibitem{Sim} B.~Simon, {\em
Functional integration and quantum physics}. Pure and Applied Mathematics, 86. Academic Press, Inc. [Harcourt Brace Jovanovich, Publishers], New York-London, 1979.

 \bibitem{Van} A. E. van de Ven, {\em Index-free heat kernel coefficients}, Classical
Quantum Gravity 15 (1998), 2311--2344

\bibitem {Vass3}D. Vassilevich, \textquotedblleft Heat kernel expansion:
user's manual", \textit{Physics Reports } \textbf{388 }(2003) 279-360.
\end{thebibliography}
\end{document}